\documentstyle[11pt,aaspp4]{article}
\def\kms{\ifmmode {\rm km\ s}^{-1} \else km s$^{-1}$\fi}
\def\ms{\ifmmode M_{\odot} \else $M_{\odot}$\fi}
\def\Lsun{\ifmmode L_{\odot} \else $L_{\odot}$\fi}
\def\Rs{\ifmmode R_{\rm S} \else $R_{\rm S}$\fi}
\def\qo{\ifmmode q_{\rm o} \else $q_{\rm o}$\fi}
\def\Ho{\ifmmode H_{\rm o} \else $H_{\rm o}$\fi}
\def\ho{\ifmmode h_{\rm o} \else $h_{\rm o}$\fi}
\def\ltsim{\raisebox{-.5ex}{$\;\stackrel{<}{\sim}\;$}}
\def\gtsim{\raisebox{-.5ex}{$\;\stackrel{>}{\sim}\;$}}
\def\vFWHM{\ifmmode V_{\mbox{\tiny FWHM}} \else
            $V_{\mbox{\tiny FWHM}}$\fi}
\def\Hbeta{\ifmmode {\rm H}\beta \else H$\beta$\fi}
\def\deg{^{\rm o}}

\lefthead{Peterson and Wandel}
\righthead{Black Holes in AGNs}

\begin{document}
\title{Evidence for Supermassive Black Holes in Active Galactic Nuclei
from Emission-Line Reverberation}

\author{
Bradley M. Peterson\altaffilmark{1}
and Amri Wandel\altaffilmark{2}
}
\altaffiltext{1}
           {Department of Astronomy, The Ohio State University,
    	140 West 18th Avenue, Columbus, OH 43210-1173\\
	Email: peterson@astronomy.ohio-state.edu}
\altaffiltext{2}
	{Racah Institute, The Hebrew University, Jerusalem 91904, ISRAEL\\
	Email: amri@frodo.fiz.huji.ac.il}

\begin{abstract}
Enission-line variability data for Seyfert 1 galaxies 
provide strong evidence for the
existence of supermassive black holes in the nuclei of these galaxies, 
and that the line-emitting gas is moving 
in the gravitational potential of that black hole.
The time-delayed response of the
emission lines to continuum variations is used to infer the
size of the line-emitting region, which is then
combined with measurements of the Doppler widths of the
variable line components to estimate a virial mass. In the case of
the best-studied galaxy, NGC~5548, various
emission lines spanning an order of magnitude
in distance from the central source show the expected
$V \propto r^{-1/2}$ 
correlation between distance and line width, and are thus
consistent with a single value for the mass. Two other Seyfert galaxies,
NGC~7469 and 3C~390.3, show a similar relationship.
We compute the ratio
of luminosity to mass for these three objects and
the narrow-line Seyfert 1 galaxy NGC~4051 and find that
that the gravitational force on the line-emitting gas
is much stronger than radiation pressure.
These results strongly support the paradigm of
gravitationally bound broad emission-line region clouds.
\end{abstract}

 
\setcounter{footnote}{0}
\section{Introduction}

Since the earliest days of research on quasars and
active galactic nuclei (AGNs), supermassive
black holes (SBHs) have been considered to be the
most likely driving power of the activity
in these sources. Indeed, compact dark masses, probably SBHs, 
have been detected in the cores of many normal galaxies using 
stellar dynamics (Kormendy \& Richstone 1995).
In the case of AGNs, however, direct detection of nuclear SBHs
through stellar-dynamical methods
has not been achieved on account of technical difficulties 
arising from the large  surface brightness
contrast between the stellar component of the galaxy and
the AGN itself on arcsecond and smaller angular scales.
Nevertheless, there is good evidence for SBHs in
AGNs. The strongest
evidence to date has been the detection of Keplerian motions
of megamaser sources in the Seyfert galaxy NGC~4258
(Miyoshi et al.\ 1995). 
Also, asymmetric Fe\,K$\alpha$ emission in the X-ray spectra of AGNs
(e.g., Tanaka et al.\ 1995) 
may show the relativistic signature of an SBH, but 
this remains somewhat speculative as 
Fe\,K$\alpha$ reverberation signatures have not yet
been observed (e.g., Reynolds 2000).

Perhaps the best method for measurement of the 
central mass in most AGNs is emission-line ``reverberation mapping'' 
which allows direct measurement of the size of the
broad-line region (BLR) in these objects (Blandford \& McKee 1982;
Peterson \& Netzer 1998).
By careful monitoring of the variability of the 
UV and optical continuum and the light travel-time delayed
response of the emission lines, one can determine the
size of the line-emitting region, or more precisely,
the distance from the central source at which the
broad emission-line response is most significant.
The broad emission lines are resolved in wavelength,
and their widths reflect the Doppler motion of
line-emitting gas.
Assuming the velocities of the line-emitting gas are controlled
by gravity, the combination of the size estimate (from 
reverberation mapping) and velocity (from the line widths)
can be used to estimate the gravitational potential in which the gas 
moves, and thus deduce central black hole masses by the virial theorem
$M \approx r \sigma^2/G$, where the line 
velocity width $\sigma$ is typically several thousands of
kilometers per second and the size of the emission-line region
$r$ is typically a few to tens of light-days.

SBH masses have been estimated using this method for about 
three dozen AGNs (Wandel, Peterson, \& Malkan 1999;
Kaspi et al.\ 2000).
However, whether or not the broad emission-line widths 
actually reflect virial motion remains an open question
and is thus a key issue in AGN mass measurement
(Richstone et al. 1998).
While the relative response times for the wings of the line
profiles reveal no strong signature of
bulk outflow, there are still viable models
with non-gravitationally driven cloud motions. However, if
the kinematics of the BLR can be proven to be gravitationally
dominated, then the reverberation results 
provide an even more definitive demonstration
of the existence of SBHs than megamaser kinematics because
the BLR is more than two orders of magnitude closer to the
central source than the megamaser sources; the inferred
SBH mass would dominate the stellar mass within the BLR 
even for the highest possible stellar densities.

Whether or not reverberation-based masses can be trusted
is an important issue on account of a possible discrepancy
between reverberation-based AGN SBH masses and
stellar-dynamical normal-galaxy SBH masses.
In the case of normal galaxies, the
SBH mass appears to correlate with the galactic
bulge luminosity, with the SBH to spheriodal bulge 
mass ratio $M_{\rm BH}/M_{\rm bulge} = 0.006$ 
(Magorrian et al.\ 1998; Richstone et al.\ 1998). 
In the case of AGNs, however, a significantly lower 
value of $M_{\rm BH}/M_{\rm bulge}$ has been found (Wandel 1999), 
indicating either a real difference between
active and normal galaxies, or that one or both of the methods of
mass determination is somehow biased. We note, however,
that other authors find lower values for non-active galaxies,
$M_{\rm BH}/M_{\rm bulge} = 0.002$--$0.003$ (Ho 1999;
Kormendy \& Ho 2000). This lower value, 
combined with the uncertainties in AGN virial masses (discussed below), 
may explain most of the discrepancy between the
black-hole masses derived by stellar-dynamical and reverberation 
methods (Wandel 2000).

Here we expand on our argument (Peterson \& Wandel 1999)
that the broad emission-line
variability data demonstrate that
the BLR kinematics are Keplerian, i.e., that the
emission-line cloud velocities are dominated by
a central mass of order of $ 10^{7-8}$\,\ms\
within the inner few light days. This seems to hold for
each of the three best-studied Seyfert 1 galaxies. We believe that
this strongly supports the hypothesis that
SBHs reside in the nuclei of active galaxies,
and underscores the importance of the reverberation
method for determination of the masses of SBHs in AGNs.

\section{The Keplerian Size-Velocity Relationship}

Measurement of the central mass using the virial mass estimate from
emission lines requires a demonstration that the kinematics 
of the line-emitting gas are gravitationally dominated.
A correlation between the broad-line width and emitting-region size 
of the form $\sigma \propto r^{-1/2}$ is consistent with a wide variety of
gravitationally dominated kinematics. It thus provides good evidence
for such a dynamical scenario, although alternative pictures which
contrive to produce a similar result may not be ruled out
definitively.

In measuring the widths of the emission lines, it is important to
include in the measurement only the part of the emission line
that is actually varying. This can be difficult on account
of contamination of the broad lines by emission from the 
narrow-line region or other non-variable (or slowly varying) components of
the line and, in some cases, contamination from other broad lines.
We circumvent this problem by using the numerous spectra obtained 
in the reverberation experiments to compute mean and
root-mean-square (rms)
spectra, and we measure the width of the emission features in
the rms spectrum. While in many cases measurement of the width
of each line from the mean spectrum gives
a similar result (Kaspi et al.\ 2000),
our procedure ensures that the emission features in
the rms spectrum accurately represent the parts of the emission
line that are varying, and for which the time delays
are measured (Peterson et al.\ 1998). 
In each rms spectrum, we determined
the full-width at half-maximum (FWHM) of each measurable line,
with a range of uncertainty estimated by the highest and lowest
plausible settings of the underlying continuum. 

In Fig.\ 1, we show the width of the line in the rms spectrum plotted
as a function of the distance from the central source 
(upper horizontal axis) measured by the emission-line lag $\tau$ (lower
horizontal axis)  for various broad emission lines in three
different Seyfert 1 galaxies, NGC 5548, NGC 7469, and 3C 390.3.
All of the data used here are publicly available on the
International AGN Watch 
website\footnote{{\sf http://www.astronomy.ohio-state.edu/$\sim$agnwatch/}}.
In the case of the best-studied galaxy, NGC 5548,
lines that are strongest in highly ionized gas
(e.g., He\,{\sc ii}\,$\lambda1640$ and  He\,{\sc ii}\,$\lambda4686$)
have the shortest response times (a few days) and the largest
Doppler widths ($\vFWHM \gtsim 8000$\,\kms), and lines that
are more prominent in less highly ionized gas (e.g.,
H$\beta\,\lambda4861$ and C\,{\sc iii}]\,$\lambda1909$) have
longer time delays (more than 10 days) and narrower widths
($\vFWHM \ltsim 7000$\,\kms).

Each data point in Fig.\ 1 provides an independent measurement of
the virial mass in each respective galaxy,
$M = f r_{\rm BLR} \sigma_{\rm rms}^2/G$, where
$r_{\rm BLR} = c \tau$, and we take
$\sigma_{\rm rms} = \sqrt{3}\vFWHM/2$ (Netzer 1990).
The factor $f$ depends on the details of the geometry, kinematics, and
orientation of the BLR and is expected to be of order
unity. 
Within the measurement uncertainties,
all the emission lines yield consistent values for
the central mass.  A weighted fit to the relationship 
$\log (\vFWHM) = a + b \log \tau $ yields
$b=-0.44\pm0.05$ for the case of NGC 5548, consistent with the expected value
$b=-1/2$, and a reduced $\chi^2_{\nu}= 3.30$ 
(compared with $\chi^2_{\nu} = 3.58$ for a
forced $b = -1/2$ fit, which is also shown in Fig.\ 1). From a 
weighted fit to these data, we obtain $M = 5.9 \pm 2.5 \times 10^7$\,\ms\
for the central mass for the SBH in NGC~5548. The formal
uncertainty reflects measurement uncertainties in the
time lag and line width. 

Moreover, we note that this
mass is {\it systematically} uncertain by a factor of a few,
on account of the uncertainty in the unknown factor $f$.
As an illustration, we consider the masses obtained from
two more detailed models of NGC 5548: if we use the
Wanders et al.\ (1995) model for the C\,{\sc iv}\,$\lambda1549$
emission line (line-emitting clouds in circular orbits of
random inclination, illuminated by a biconical beam), the
rms line width and transfer function centroid match the
observations best for a central mass of $M = 2.4 \times 10^{7}\,\ms.$
Alternatively, if we try to match the \Hbeta\ rms line width
and transfer function centroid with a disk model and reasonable photoionization
model parameters (Ferland et al.\ 1992), then masses in the range
$(1.4 $--$ 6.0) \times 10^{7}\,\ms$ are obtained as the inclination
is varied from $90\deg$ to $\sim30\deg$.

Although NGC~5548 has by far the most detailed and highest-quality 
reverberation data, it is not the only AGN that
shows the Keplerian relationship.
We are able to derive reliable reverberation sizes and rms profiles 
for multiple emission lines in at least two more Seyfert galaxies,
NGC~7469 (Wanders et al.\ 1997; Collier et al.\ 1998; Kriss
et al.\ 2000) 
and 3C~390.3 (Dietrich et al.\ 1998; O'Brien et al.\ 1998),
also shown in Fig.\ 1.
In both cases the virial mass values calculated from each line are
consistent with the same central mass 
(for NGC~7469, $M=8.4\times 10^6\,\ms$,
$\sim7$ times smaller than NGC~5548, and for 3C~390.3, 
$M=3.2\times 10^8\,\ms$, $\sim5$ times larger than NGC~5548).
Together these three examples span a range of almost two decades in 
virial mass, a factor of three in line width and of five in BLR size, 
suggesting that Keplerian radial velocity profiles are common in
a wide range of AGNs.

\section{Complications and Caveats}

These results are consistent with a surprisingly
simple description of the BLR. At least to first order,
the BLR gas appears to orbit the central source,
possibly in a flattened geometry related to the
accretion-disk structure, although ellipsoidal or
spherical geometries cannot be excluded.
The fact that the highest
ionization lines have the smallest response times 
and the largest Doppler widths indicates
that the BLR has an ionization-stratified structure.
The \Hbeta\ response time in NGC~5548 is known to
depend on the mean luminosity of the central source
in the sense that the emission-line lag is longer
when the continuum source is brighter
(Peterson et al.\ 1999).
We also see that when the emission-line lag is longer,
the line width is smaller, apparently because we
are seeing enhanced response from line-emitting
gas that is farther away from the central source.

While some of the important general characteristics of
AGN emission lines and their variability are accounted for
with such a simple explanation, it
is nevertheless clear that the actual situation is more complex.
A responsivity map of each line would presumably reveal an
extended and complex geometry, similar to what is
seen in the spatially extended narrow-line regions.

The most important difficulty is that accurate measurement of
the SBH mass depends on knowing the detailed geometry and orientation of
the BLR, i.e., determining the factor $f$.
For example, if the BLR is a flattened disk, the virial mass
we deduce can be off by a factor of a few on account of
the unknown inclination. Is there in fact evidence that
geometrical effects might be important? We believe that there
is. While the fits to the data above are consistent with
the virial interpretation, there is significant residual scatter.
The residual scatter in the relationship may be 
at least in part ascribable to somewhat
different geometries of the line-emitting region
at different radii, i.e., different values of $f$;
indeed, it has been often argued that the high- and
low-ionization lines arise in very different regions
(Collin-Souffrin et al.\ 1988).
A notable example in this regard is the narrow-line
Seyfert~1 galaxy NGC~4051, in which the hydrogen
Balmer lines might arise in a low-inclination (nearly face-on)
disk but the high-ionization 
lines may arise in an outflowing wind above and below the disk
(Peterson et al.\ 2000).
At the other extreme, the Balmer-line profiles  
of 3C 390.3 are double peaked, which is the expected profile
of a high-inclination disk. In any case, 3C 390.3 may have
a somewhat atypical BLR structure, however, as the usual pattern of
increasing lag with decreasing ionization level seems to be
reversed in this object (i.e., C\,{\sc iv}\,$\lambda1549$ has
a larger lag and narrower Doppler width than \Hbeta,
in contrast to the trend seen in every other object that
has been studied).
Interpretation of double-peaked line profiles
rather problematic; in NGC~5548 and Akn~120, for example,
the Balmer lines are sometimes double-peaked, but single-peaked
or multiple-peaked at other times
(Peterson, Pogge, \& Wanders 1999).

In spite of these uncertainties, the data at hand seem to rule out
models that predict a distance--radial-velocity relationship
that is significantly different from the Keplerian case.
For example, in most outflow models, a wind or radiation pressure-driven
acceleration is assumed. In this case the velocity should increase
with radius, or at least remain constant.
A constant velocity width over the radial range sampled by the 
data of the reverberation-mapped emission lines can be ruled out at the
confidence levels of $\gtsim99.9$\%,
$\gtsim99$\%, and $\gtsim95$\% for  NGC~5548, 3C~390.3, and 
NGC~7469, respectively.

\section{Mass vs.\ Luminosity }

In order to determine whether the BLR could be radiatively accelerated, we
estimate the ratio of gravitational to radiation pressure.
This can be formulated in terms of the well-known Eddington 
limit, the luminosity above which the radiation pressure exceeds the
gravitational force on an ionized gas, $L_{\rm Edd}=
4\pi GMm_{\rm H}/\sigma_{\rm T} = 3 \times 10^{11}\,L_{\odot}\,
(M/10^7 M_{\odot})$.  We can further define the
Eddington ratio $F_{\rm rad}/F_{\rm grav} = L/L_{\rm Edd}$.  We have seen
how reverberation data on the broad lines can give us a reliable estimate
for the central mass. If in addition the total continuum luminosity can be
estimated, it is possible to calculate the Eddington ratio.
We note, of course, that the Eddington limit applies to the case
of spherical accretion, and that super-Eddington accretion rates
can be achieved in other geometries. The calculation is nevertheless
interesting in that we can use it to argue against the case 
of radiatively driven spherical outflow. 

Most of the AGN energy is radiated in the far-UV band, in the so
called ``UV bump'' or ``big blue bump,'' of which only the long-wavelength
tail is directly observable.  A very crude estimate of the UV-bump flux
can be made by multiplying
the energy flux in the observable spectrum by a constant bolometric
correction factor, 
but this does not take into account the differences between the
spectra of different AGNs (for example, more luminous objects appear to
have flatter UV spectra, possibly indicating a more prominent UV bump). 
A presumably more accurate estimate 
can be made by inferring the ionizing luminosity from the
broad emission lines, since the broad lines are powered by the
ionizing continuum.
By comparing the BLR size derived from reverberation data for some 
particular emission
line (we use H$\beta$) to the $R \propto L^{1/2}$ scaling that is 
predicted by naive photoionization theory,
one can estimate the ionizing
continuum luminosity from the central source
(Wandel et al.\ 1999). From the definition of the ionization parameter
$U = L_{\rm ion}/ 4\pi r^2 \bar{E} n c$, where $\bar {E}$ is the average
photoionizing photon energy, and $n$ the mean particle density in
the line-emitting gas,  the  BLR radius $r$ can be calculated by
using an estimate of $L$. Inverting this method and 
measuring $r$ independently by reverberation methods, we can derive
an estimate for the ionizing luminosity 
\begin{equation}
L_{\rm ion} \approx 6\times 10^{43} \left ( {E\over{\rm 1\,Ryd}}\right ) 
\left( {n\over 10^{10}\,{\rm cm}^{-3}} \right) 
\left ({\tau\over 10 {\rm \,days}}\right )^2 {\rm ergs ~s}^{-1},
\end{equation} 
where $\tau$ is the time lag for the H$\beta$ emission-line response,
and the normalization has been determined from the Seyfert 1 sample
of Wandel et al.\ (1999).
Since the ionizing luminosity
represents a significant fraction of the bolometric luminosity, together
with the virial mass this implies that $L/L_{\rm Edd} \propto L_{\rm
ion}/M_{\rm vir}$. 

In Fig.\ 2, we show both $L_{\rm ion}$ and the optical luminosity
$L_{\rm vis}$ versus
$M_{\rm vir}$ for the three objects discussed here.
In order to extend the range in mass and spectral properties,
we have also included new results
(Peterson et al.\ 2000)
on the narrow-line Seyfert 1 galaxy NGC~4051.
With the exception of NGC~7469, the visual luminosity 
of these objects is 
of order $\sim0.001 L_{\rm Edd}$ and $L_{\rm ion}/L_{\rm Edd}\approx 0.01$. 
Typically $L_{\rm ion}$ is larger than $L_{\rm vis}$ by a factor of 
10--30 (Wandel et al.\ 1999).
NGC 7469 seems to have an exceptionally short lag relative to
its visual luminosity,  which gives $L_{\rm ion}$ {\em smaller}
than $L_{\rm vis}$ and a relatively high value of 
$L_{\rm vis}/L_{\rm Edd}\approx 0.03$.
For NGC~4051, $L_{\rm ion}\approx 0.1$, larger than $ L_{\rm vis}$  
by a factor of 100 (presumably indicating the harder
UV spectrum and soft X-ray excess typical of narrow-line Seyfert 1s).
We note that the ratio $L_{\rm ion}/L_{\rm Edd}$ 
could be even smaller for a low-inclination flattened 
geometry, in which case our derived values
of $L/L_{\rm Edd}$ are lower limits.

The important things to note are that 
(1) in any case $L/L_{\rm Edd}$ is small enough for gravity to dominate
even after a bolometric correction and allowing for the 
uncertainty in the reverberation mass estimate, and 
(2) not surprisingly, the relationship between $L_{\rm ion}$
and $M_{\rm vir}$ has less scatter than the relationship between
$L_{\rm vis}$ and $M_{\rm vir}$.

\medskip

\acknowledgements{BMP is grateful for support of this work
by the National Science Foundation and NASA through grants
AST--9420080 and NAG5--8397, respectively,
to The Ohio State University. AW wishes to
acknowledge the hospitality of the Department of Physics and
Astronomy at UCLA  and Stanford University during this work.
We thank Dr.\ Alex Filippenko for a predictably helpful and
insightful referee's report on this paper.}

\clearpage

\begin{figure}
\caption{Line width in the
rms spectrum plotted as a function of the distance from the central source
(upper horizontal axis) as measured by the emission-line lag 
(lower horizontal axis) for various broad emission lines in 
NGC~7469, NGC~5548, and 3C 390.3. The dashed lines are 
best fits of each set of data to the relationship
$\log \vFWHM = a + b\log c\tau$, and the best-fit
slopes are $b=-0.61\pm0.35$, $-0.44\pm0.05$, and 
$0.41\pm0.15$ for the three galaxies, respectively.
The solid line shows the best fit to each set of data for fixed $b=-1/2$,
yielding virial masses of 
$8.4\times 10^6\,\ms$, 
$5.9\times 10^7\,\ms$, and 
$3.2\times 10^8\,\ms$ for the three respective galaxies.}
\end{figure}

\begin{figure}
\caption{The luminosity versus virial mass relationship for four AGNs.
Squares with error bars denote $L_{\rm ion}$, with virial
masses based on all the lines shown in Fig.\ 1. The filled
circles represent $L_{\rm vis}$, with virial masses based on
Balmer lines only, as by Wandel, Peterson, \& Malkan (1999)
and Kaspi et al.\ (2000),  with the
error bars omitted for clarity.
The lines show constant values of the Eddington ratio $L/L_{\rm Edd}$
between 0.001 and 1.}
\end{figure}

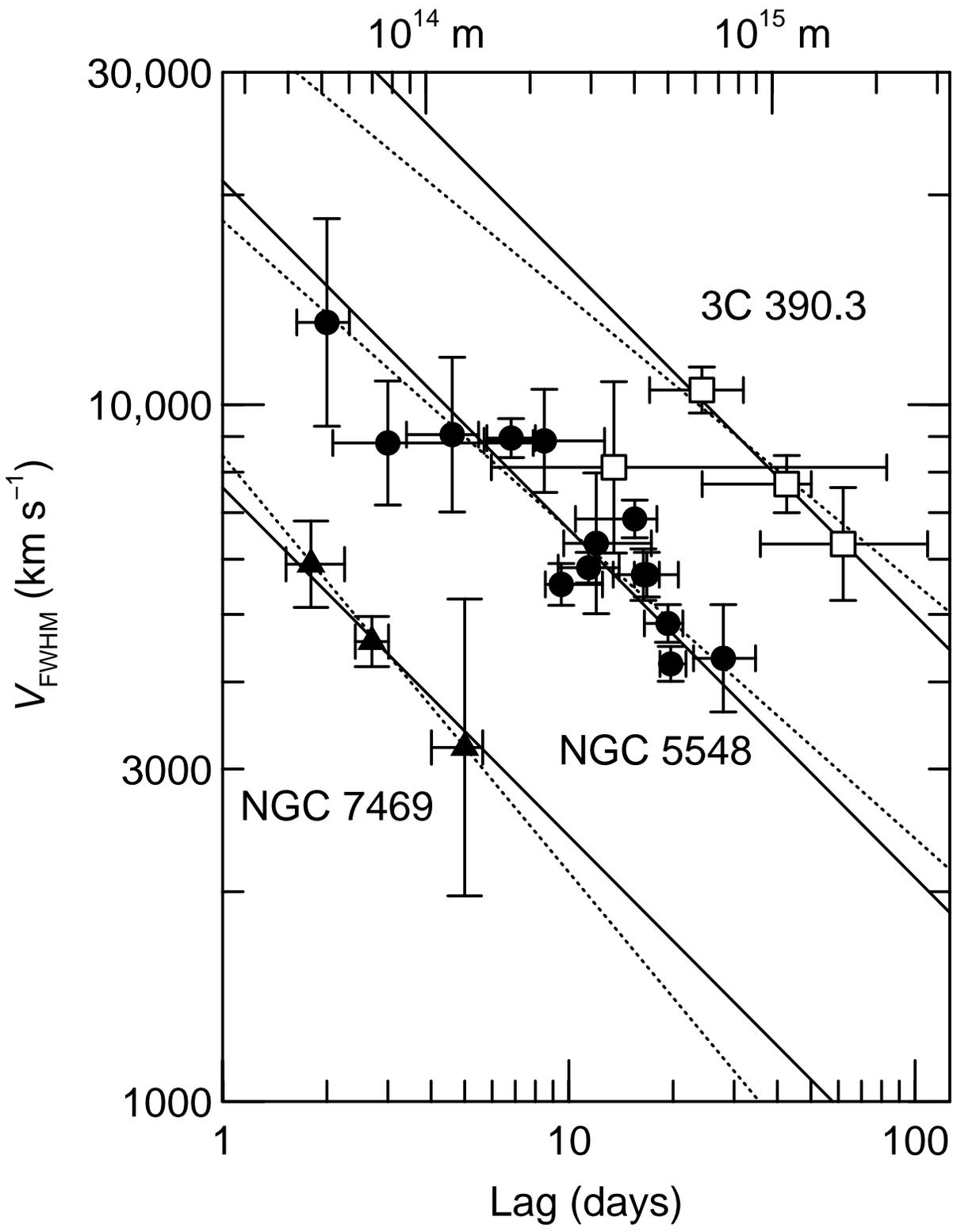
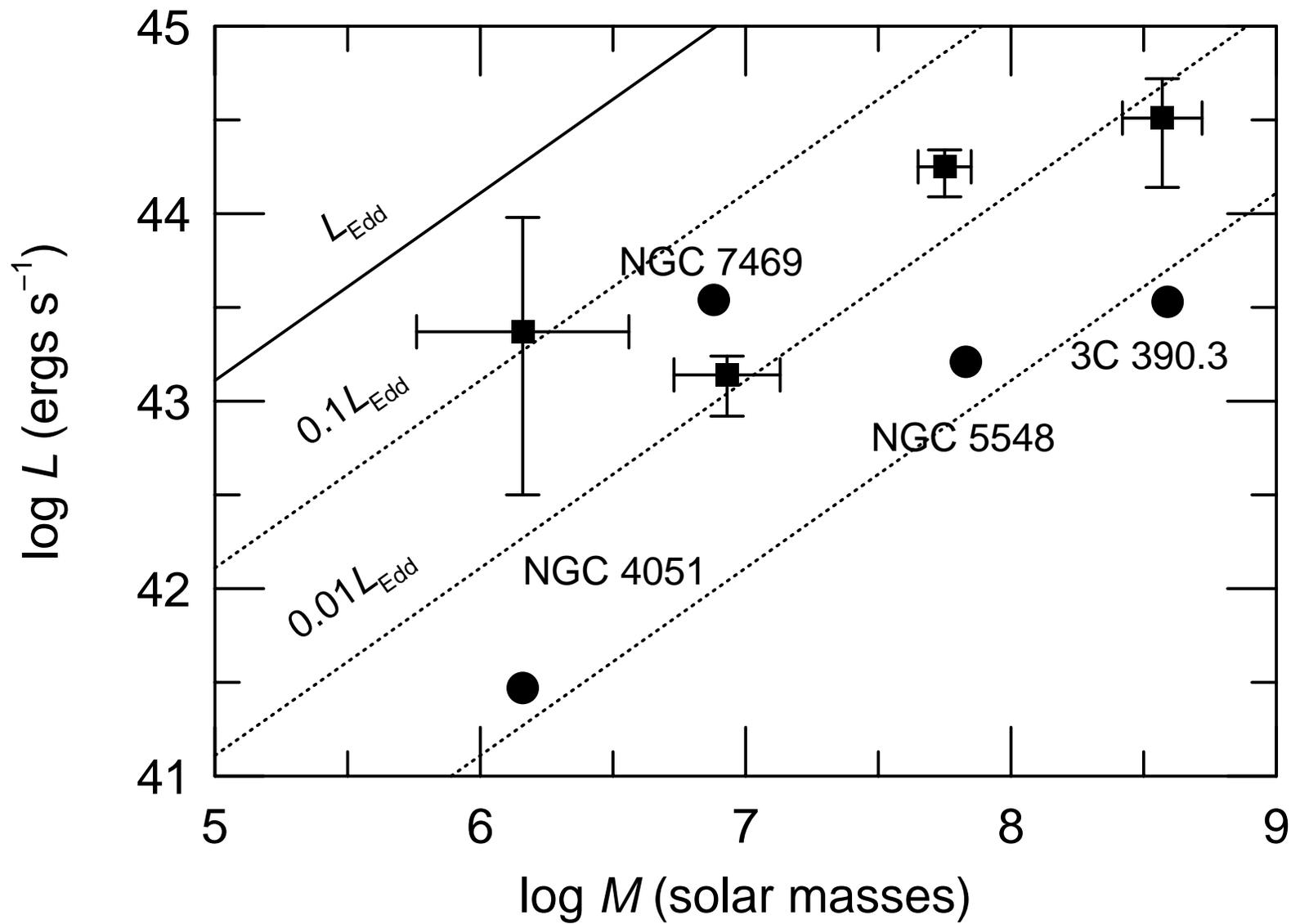
\end{document}